\documentclass[aps,pre,floatfix,twocolumn,groupedaddress]{revtex4-1}
\usepackage[dvips]{graphicx}
\usepackage{amsmath}
\usepackage{verbatim}
\usepackage{color}
\newcommand{\la}{\left\langle}
\newcommand{\ra}{\right\rangle}
\newcommand{\br}{{\bf r}}

\newcommand{\PRL}{Phys.~Rev.~Lett.~}
\newcommand{\PR}{Phys.~Rev.~}
\newcommand{\JCP}{J.~Chem.~Phys.~}

\newcommand{\JPCM}{J.~Phys.: Condens.~Matter~}

\begin{document}

\title{Influence of Solvent Quality on Depletion Potentials in Colloid-Polymer Mixtures}

\author{Alan R. Denton and Wyatt J. Davis}
\email[]{alan.denton@ndsu.edu}
\affiliation{Department of Physics, North Dakota State University,
Fargo, ND 58108-6050, USA}

\begin{abstract}
As first explained by the classic Asakura-Oosawa (AO) model, effective attractive forces
between colloidal particles induced by depletion of nonadsorbing polymers can drive demixing 
of colloid-polymer mixtures into colloid-rich and colloid-poor phases, with practical
relevance for purification of water, stability of foods and pharmaceuticals, and 
macromolecular crowding in biological cells.
By idealizing polymer coils as effective penetrable spheres, the AO model qualitatively captures
the influence of polymer depletion on thermodynamic phase behavior of colloidal suspensions.
In previous work, we extended the AO model to incorporate aspherical polymer conformations
and showed that fluctuating shapes of random-walk coils can significantly modify
depletion potentials [W.~K.~Lim and A.~R.~Denton, {\it Soft Matter} {\bf 12}, 2247 (2016);
{\it J.~Chem.~Phys.}~{\bf 144}, 024904 (2016)]. We further demonstrated that the shapes 
of polymers in crowded environments depend sensitively on solvent quality
[W.~J.~Davis and A.~R.~Denton, {\it J.~Chem.~Phys.}~{\bf 149}, 124901 (2018)].
Here we apply Monte Carlo simulation to analyze the influence of solvent quality 
on depletion potentials in mixtures of hard-sphere colloids and nonadsorbing 
polymer coils, modeled as ellipsoids whose principal radii fluctuate according to 
random-walk statistics. We consider both self-avoiding and non-self-avoiding 
random walks, corresponding to polymers in good and theta solvents, respectively. 
Our simulation results demonstrate that depletion of polymers of equal molecular weight 
induces much stronger attraction between colloids in good solvents than in 
theta solvents and confirm that depletion interactions are significantly 
influenced by aspherical polymer conformations.
\end{abstract}

\maketitle
\newpage

%%%%%%%%%%%%%%%%%%%%%%%%%%%%%%%%%%%%%%%%%%%%%%%%%%%%%%%%%%%%%%%%%%%%%%%%%%%
%%%%%%%                       BODY OF TEXT
%%%%%%%%%%%%%%%%%%%%%%%%%%%%%%%%%%%%%%%%%%%%%%%%%%%%%%%%%%%%%%%%%%%%%%%%%%%

\section{Introduction}\label{introduction}

In the field of condensed matter physics, the Asakura-Oosawa (AO) model of
colloid-polymer mixtures~\cite{asakura1954,binder2014} has a status akin 
perhaps to the van der Waals model of fluids, the Einstein model of solids, 
the Ising model of magnets, and the primitive model of electrolytes. The AO model, 
inspired by deep physical insight, first identified polymer depletion as the 
basic mechanism underlying effective attraction between colloidal particles 
induced by nonadsorbing polymers. In soft, fragile materials, depletion-induced 
interactions~\cite{vrij1976,pusey1991,lekkerkerker-tuinier2011,fuchs2002,fleer-tuinier2008}
between mesoscopic particles typically compare in magnitude to thermal energies 
and thus can strongly influence self-assembly and thermodynamic phase stability.
In this way, the AO model qualitatively explains observed phase behavior 
of colloid-polymer mixtures, in particular, demixing into colloid-rich and 
colloid-poor bulk phases.

In many practical applications, such as in stabilizing foods~\cite{tolstoguzov1991,
deKruif-tuinier2001} and pharmaceuticals against coagulation or preventing
the aggregation of proteins~\cite{zukoski2001,stradner2007}, it is important 
to minimize depletion-induced attraction. In other applications, such as in 
purifying water by promoting flocculation of colloidal impurities~\cite{norde2011}
or in guiding the self-assembly of virus particles~\cite{dogic2004,li2013}, 
amplifying the effects of polymer depletion is instead desirable.
Depletion also contributes to macromolecular crowding and segregation of
biopolymers within biological cells~\cite{minton1981,minton2000,minton2001,minton2005,
richter2007,richter2008,elcock2010,hancock2012}.

In its original form, the AO model depicts polymer coils as effective spheres, 
of fixed size defined by the radius of gyration, that are mutually penetrable,
but impenetrable to colloidal particles due to excluded-volume interactions. 
The model reveals that depletion of polymer from the space between hard colloidal 
surfaces creates an imbalance in polymer concentration, and thus in
osmotic pressure, that drives effective attraction between colloids.
Equivalently, configurations in which excluded-volume shells of neighboring colloids 
overlap maximize the free volume available to polymer coils and thus are entropically favored.

Although it captures the essence of polymer depletion, the AO model omits certain 
important aspects of real physical systems. Most obviously, by idealizing polymer
coils as effective spheres of unvarying size, the model neglects the internal 
degrees of freedom -- structure and flexibility -- of polymers in solution.
In biological systems, structure associated with folding (or misfolding) of proteins,
determines the function of such biopolymers, with relevance for many diseases.

The realization that polymers are flexible, aspherical objects predates the AO model 
by at least two decades. Kuhn argued~\cite{kuhn1934} that linear polymer coils 
in solution can be modeled as random walks with fluctuating shapes that 
approximate those of elongated, flattened ellipsoids (in their principal-axis frame).
The insight that the end-to-end path of a polymer is a physical manifestation
of a random walk has spurred many studies
of shapes of random walks~\cite{fixman1962,flory-fisk1966,flory1969,yamakawa1970,
fujita1970,solc1971,solc1973,theodorou1985,rudnick-gaspari1986,rudnick-gaspari1987,
bishop1988,sciutto1996,schaefer1999,murat-kremer1998,eurich-maass2001}.
As a vital example, the shapes of RNA, DNA, and proteins are important 
for cellular processes in the crowded environment of 
biological cells~\cite{ellis2001a,ellis2001b,vandermaarel2008,phillips2009,cheung2013,
denton-cmb2013}, translocation of polymers through narrow pores~\cite{grosberg2006,polson2013},
and packaging of DNA in viral capsids~\cite{yeomans2006}.

Depletion forces and their impact on polymer crowding and phase behavior in 
colloid-polymer mixtures have been probed by neutron scattering~\cite{tong1996,han2001,
kramer2005a,kramer2005c,longeville2009,longeville2010,richter2010},
atomic force microscopy~\cite{milling-biggs1995},
total internal reflection microscopy~\cite{leiderer1998}, 
optical trapping~\cite{yodh1998,yodh2001,dogic2015}, and 
turbidity measurements~\cite{vanduijneveldt2005,vanduijneveldt2006,vanduijneveldt2007},
to name but a few experimental methods.
Modeling studies of colloid-polymer mixtures have used scaling and mean-field 
free-volume theories~\cite{deGennes1979,doi-edwards1986,joanny-leibler-deGennes1979,
sear1997,sear2001,sear2002,denton-schmidt2002,lu-denton2011,lim-denton2014}, 
force-balance theory~\cite{walz-sharma1994},
perturbation theory~\cite{mao-cates-lekkerkerker1995,mao-cates-lekkerkerker1997},
polymer field (renormalization group) theories~\cite{eisenriegler1996,hanke1999,
eisenriegler2003,odijk2000,forsman2010,forsman2012,forsman2014,ganesan2005},
integral-equation theories~\cite{chatterjee1998a,chatterjee1998b,chatterjee1999,schweizer2002,moncho-jorda2003},
density-functional theories~\cite{leiderer1999,schmidt-fuchs2002,mukherjee2004,
forsman2008,forsman2009}, 
adsorption theories~\cite{tuinier-lekkerkerker2000,tuinier-lekkerkerker2001,tuinier-petukhov2002},
and computer simulation of molecular~\cite{meijer-frenkel1991,meijer-frenkel1994,dickman1994,
bolhuis2002,louis2002-jcp2,bolhuis2003,doxastakis2004,doxastakis2005,goldenberg2003,dima2004,cheung2005,
denesyuk2011,denesyuk-thirumalai2013a,likos2010,linhananta2012,wittung-stafshede2012}
and coarse-grained~\cite{hoppe2011,denesyuk-thirumalai2013b,lu-denton2011} polymer models.

Previous studies have investigated depletion forces induced by aspherical depletants 
(e.g., rods, ellipsoids) that are fixed in size and shape~\cite{mao-cates-lekkerkerker1997,
kamien1999,piech-walz2000,yodh2001}.
Recently, we explored polymer crowding and depletion forces in models of colloid-polymer
mixtures, with polymers modeled as fluctuating, penetrable ellipsoids, in both 
$\theta$ solvents~\cite{lim-denton2014,lim-denton-jcp2016,lim-denton-sm2016} and 
good solvents~\cite{davis-denton2018}, distinguished by whether polymer segments 
are effectively ideal (noninteracting) or nonideal, excluding volume to one another.
The purpose of the present paper is to assess the interconnected influences of 
polymer shape and solvent quality on depletion interactions in colloid-polymer mixtures.

The remainder of the paper is organized as follows. In Sec.~\ref{models}, we describe 
modeling of linear polymer coils both as random walks and as equivalent ellipsoids that 
fluctuate in size and shape. The statistics governing conformational fluctuations depend 
on whether a coil is modeled as a self-avoiding walk, appropriate for a polymer in a
good solvent, or as a non-self-avoiding random walk, corresponding to a polymer in a 
$\theta$ solvent.
In Sec.~\ref{simulations}, we outline our numerical methods, based on Monte Carlo
estimation of average polymer depletion volume, for computing the potential of mean force 
(PMF) between hard-sphere colloids induced by depletion of nonadsorbing polymer in solvents
of differing qualities. Section~\ref{results} presents results of our calculations 
of PMFs for fluctuating ellipsoidal and fixed spherical polymers in both good and 
$\theta$ solvents. Section~\ref{conclusions} summarizes our study and concludes 
with an outlook for possible future work.

\section{Models}\label{models}

As noted above, the classic Asakura-Oosawa coarse-grained model of 
colloid-polymer mixtures~\cite{asakura1954,binder2014} idealizes
nonadsorbing polymer coils as effective spheres of fixed size.
The spherical polymer approximation, while incorporating an important
length scale, ignores aspherical conformations and fluctuations in conformation,
both of which can significantly affect depletion-induced forces.
As in our earlier work on polymer crowding~\cite{lim-denton2014,
lim-denton-jcp2016,lim-denton-sm2016,davis-denton2018}, we extend the AO model 
by representing the polymers as effective ellipsoids that fluctuate in size and shape 
according to random-walk statistics. In the current study, we consider only
the colloid limit, in which the colloids are larger than the polymer coils,
such that penetration of polymer by colloids is negligible.
Although we focus here on linear homopolymers, the analysis below is easily generalized 
to other macromolecular architectures~\cite{eurich2007}.

\subsection{Polymer Coils as Random Walks}\label{polymer-rw}

The size and shape of a polymer coil composed of $N$ identical segments linked  
to form a connected chain can be characterized by a gyration tensor ${\bf T}$, 
expressed as a matrix with elements
\begin{equation}
T_{ij}=\frac{1}{N}\sum_{k=1}^N r_{ki} r_{kj}~, 
\label{gyration-tensor}
\end{equation}
where $r_{ki}$ denotes the $i^{\rm th}$ component of the position vector ${\bf r}_k$ 
of the $k^{\rm th}$ segment relative to the coil's center of mass. The gyration tensor 
relates to the moment of inertia tensor ${\bf I}$ of a rigid body 
via ${\bf T}=R_p^2{\bf 1}-{\bf I}$, where ${\bf 1}$ is the unit tensor and
\begin{equation}
R_p=\left(\frac{1}{N}\sum_{i=1}^Nr_i^2\right)^{1/2}=\sqrt{\Lambda_1+\Lambda_2+\Lambda_3}
\label{radius-gyration}
\end{equation}
is the radius of gyration of a particular coil conformation expressed in terms of
the eigenvalues of ${\bf T}$, $\Lambda_i$ ($i=1,2,3$).
Note that $R_p$ is invariant with respect to change of reference frame.
The root-mean-square (rms) radius of gyration -- a property of a polymer coil that
is experimentally measurable (e.g., via neutron or light ray scattering) -- 
relates to the eigenvalues via
\begin{equation}
R_g=\sqrt{\la R_p^2\ra}=\sqrt{\la\Lambda_1+\Lambda_2+\Lambda_3\ra}~,
\label{gyration-avg}
\end{equation}
where the angular brackets represent an ensemble average over polymer conformations.

If the ensemble average in Eq.~(\ref{gyration-avg}) is evaluated relative to
a fixed frame of reference in which the polymer coil rotates, then the average 
tensor describes a sphere, represented by a symmetric matrix with equal eigenvalues.
On the other hand, if the average is evaluated relative to a reference frame 
that rotates with the principal axes of the coil, and if coordinate axis labelling  
preserves the order of the eigenvalue magnitudes ($\Lambda_1>\Lambda_2>\Lambda_3$),
then the average tensor describes an anisotropic object, represented by
an asymmetric matrix~\cite{rudnick-gaspari1986,rudnick-gaspari1987}.
In other words, the average shape of a fluctuating random walk is spherical when 
viewed from a fixed frame of reference, but distinctly aspherical -- shaped
like an elongated, flattened bean -- when viewed from the principal-axis 
frame~\cite{kuhn1934,solc1971,solc1973}.

\subsection{Polymer Coils as Fluctuating Ellipsoids}\label{polymer-ellipsoid}

The shape of a polymer coil in the principal-axis frame of reference can be 
fit by a general ellipsoid with principal radii proportional to the 
square-roots of the respective eigenvalues of the gyration tensor. 
In Cartesian $(x,y,z)$ coordinates, the ellipsoid surface is described by
\begin{equation}
\frac{x^2}{\Lambda_1}+\frac{y^2}{\Lambda_2}+\frac{z^2}{\Lambda_3}=3~.
\label{ellipsoid}
\end{equation}
%Added:
%\textcolor{red}{
Note that the actual shape of the coil is not necessarily ellipsoidal, but is 
approximated by an ellipsoid whose principal radii have the same proportions
as those derived from the gyration tensor.
%}
%
For a freely-jointed polymer coil of $N$ segments of Kuhn length $l$, corresponding to 
an ideal non-self-avoiding random walk (RW)~\cite{murat-kremer1998}, 
modeling conformations of a linear polymer dispersed in a $\theta$ solvent, the shape 
probability distribution is accurately fit by the analytical form~\cite{eurich-maass2001}
\begin{equation}
P(\lambda_1,\lambda_2,\lambda_3) = f_1(\lambda_1)f_2(\lambda_2)f_3(\lambda_3)~,
\label{Plambda}
\end{equation}
where $\lambda_i\equiv\Lambda_i/(Nl^2)$ are scaled (dimensionless) eigenvalues and 
the fit functions are given by
\begin{equation}
f_i(\lambda_i) = \frac{(\alpha_i d_i)^{n_i-1}\lambda_i^{-n_i}}{2K_i}
\exp\left(-\frac{\lambda_i}{\alpha_i}-d_i^2\frac{\alpha_i}{\lambda_i}\right)~,
\label{fi-Eurich-Maass}
\end{equation}
with fit parameters, $K_i$, $\alpha_i$, $d_i$, and $n_i$ listed in Table~\ref{table1}.
The rms radius of gyration of an ideal polymer coil is $R_g=\sqrt{N/6}\,l$, 
while the principal radii of the ellipsoid representing a particular conformation
are given by 
\begin{equation}
R_i(\lambda_i)=\sqrt{18\lambda_i}\,R_g~, \quad i=1,2,3~.
\label{principal-radii}
\end{equation}
It should be noted that the factorization ansatz of Eq.~(\ref{Plambda}) is not exact,
since extensions of a random walk in orthogonal directions are not strictly independent.
Nevertheless, conformations that significantly violate the ansatz occur only rarely
for sufficiently long polymer chains.

\begin{table}
\caption{Fit parameters for probability distribution of eigenvalues of gyration tensor of a 
linear polymer coil in a $\theta$ solvent modeled as a non-self-avoiding random walk 
[Eqs.~(\ref{Plambda}), (\ref{fi-Eurich-Maass})].}
\vspace*{0.2cm}
\centering
\begin{tabular}{|c|c|c|c|c|}
\hline
eigenvalue $i$ & $K_i$ & $\alpha_i$ & $d_i$ & $n_i$ \\
\hline
\hline
1&0.094551&0.08065&1.096&1/2\\
\hline
2&0.0144146&0.01813&1.998&5/2\\
\hline
3&0.0052767&0.006031&2.684&4\\
\hline
\end{tabular}
\label{table1}
\end{table}

For self-avoiding walks (SAW), modeling conformations of linear polymers dispersed 
in a good solvent, whose segments exclude volume to one 
another~\cite{flory1969,deGennes1979,doi-edwards1986}, 
the rms radius of gyration is related to the segment 
number via $R_g=CN^{\nu}l$, with Flory exponent $\nu=0.588$ and amplitude 
$C=0.44108$~\cite{sciutto1996}. (For an ideal polymer in a $\theta$ solvent, 
$\nu=1/2$ and $C=1/\sqrt{6}=0.40825$.) 
As a common example, for polystyrene, good solvents are benzene, toluene, and chloroform,
while typical $\theta$ solvents are cyclohexane and decalin, depending on temperature.

Since the gyration tensor eigenvalues 
vary as $N^{2\nu}$, the scaled eigenvalues are now defined as 
$\lambda_i\equiv\Lambda_i/(N^{2\nu}l^2)$ and are related to the principal radii via
\begin{equation}
R_i(\lambda_i)=\frac{R_g}{C}\sqrt{3\lambda_i}=3.9269\sqrt{\lambda_i}\,R_g~.
\label{Ri}
\end{equation}
The shape probability distribution is accurately fit by the same factorized function 
as for RW polymers [Eq.~(\ref{Plambda})], but with fit functions of the form
\begin{equation}
f_i(\lambda_i)=a_i \lambda_i^{b_i} \exp(-c_i\lambda_i)~.
\label{fi-Sciutto}
\end{equation}
The fit parameters $a_i$, $b_i$, and $c_i$ are tabulated in Table~\ref{table2}.
It should be emphasized that the eigenvalue distributions [Eqs.~(\ref{fi-Eurich-Maass}) 
and (\ref{fi-Sciutto})] are fits to statistics from molecular simulations of 
linear polymer chains~\cite{murat-kremer1998,eurich-maass2001,sciutto1996} and
reflect considerable fluctuations in polymer size and shape.

\begin{table}
\caption{Fit parameters for probability distribution of eigenvalues of gyration tensor  
of a linear polymer coil in a good solvent modeled as a self-avoiding walk 
[Eqs.~(\ref{Plambda}), (\ref{fi-Sciutto})].}
\vspace*{0.2cm}
\centering
\begin{tabular}{|c|c|c|c|}
\hline
eigenvalue $i$ & $a_i$ & $b_i$ & $c_i$ \\
\hline
\hline
1& 11847.9 & 2.35505 & 22.3563 \\
\hline
2& 1.11669$\times 10^9$ & 3.71698 & 148.715 \\
\hline
3& 1.06899$\times 10^{14}$ & 4.84822 & 543.619 \\
\hline
\end{tabular}
\label{table2}
\end{table}

The average shape of a polymer coil can be quantified by an 
asphericity parameter~\cite{rudnick-gaspari1986,rudnick-gaspari1987}, 
\begin{equation}
{\cal A}\equiv 1-3\frac{\la\lambda_1\lambda_2+\lambda_1\lambda_3+\lambda_2\lambda_3\ra}
{\la(\lambda_1+\lambda_2+\lambda_3)^2\ra}~,
\label{asphericity}
\end{equation}
defined such that a perfect sphere, with all eigenvalues equal, has ${\cal A}=0$, while a 
needle-like object has ${\cal A}\simeq 1$. Interestingly, for both RW and SAW coils, 
${\cal A}\simeq 0.54$ in uncrowded environments~\cite{davis-denton2018}.

\begin{figure}[t!]
\begin{center}
\includegraphics[width=\columnwidth]{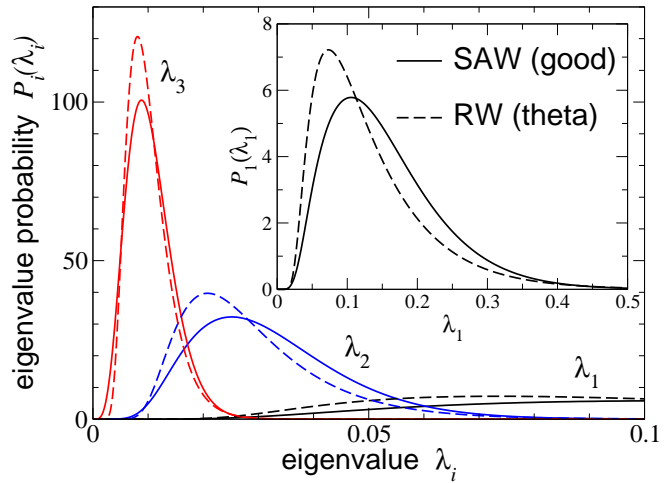}
\end{center}
\vspace*{-0.2cm}
\caption{Probability distributions $P_i(\lambda_i)$ [Eqs.~(\ref{P1})-(\ref{P3})]
of scaled eigenvalues $\lambda_i$ ($i=1, 2, 3$) of gyration tensor for polymers 
in a good solvent (SAW, solid curves) and $\theta$ solvent (RW, dashed curves).
Inset: distributions for largest eigenvalue $\lambda_1$.}
\label{fig2}
\end{figure}

The probability distributions $P_i(\lambda_i)$ for the individual eigenvalues differ somewhat 
from the fit functions $f_i(\lambda_i)$ in Eqs.~(\ref{fi-Eurich-Maass}) and (\ref{fi-Sciutto}). 
Each is obtained from the parent distribution [Eq.~(\ref{Plambda})] by integrating over the 
other two eigenvalues, with limits imposed by the requirement of
eigenvalue ordering ($\lambda_1>\lambda_2>\lambda_3$):
\begin{equation}
P_1(\lambda_1)=
\int_0^{\lambda_1}d\lambda_2\, \int_0^{\lambda_2}d\lambda_3\, 
P(\lambda_1, \lambda_2, \lambda_3)~,
\label{P1}
\end{equation}
\begin{equation}
P_2(\lambda_2)=
\int_{\lambda_2}^{\infty}d\lambda_1\, \int_0^{\lambda_2}d\lambda_3\, 
P(\lambda_1, \lambda_2, \lambda_3)~,
\label{P2}
\end{equation}
\begin{equation}
P_3(\lambda_3)=
\int_{\lambda_3}^{\infty}d\lambda_1\, \int_{\lambda_3}^{\lambda_1}d\lambda_2\, 
P(\lambda_1, \lambda_2, \lambda_3)~.
\label{P3}
\end{equation}
For comparison, Fig.~\ref{fig2} shows the scaled eigenvalue distributions of polymers
in $\theta$ and good solvents.
Note that, accounting for the different scaling factors -- $N$ for RW polymers, but $N^{1.176}$
for SAW polymers -- the unscaled eigenvalues are significantly larger for a SAW polymer in
a good solvent than for a RW polymer in a $\theta$ solvent, reflecting the more
extended conformations of polymers with excluded-volume interactions.

\subsection{Colloid-Polymer Mixtures}\label{cp-mixtures}

\begin{figure}[t!]
\includegraphics[width=\columnwidth]{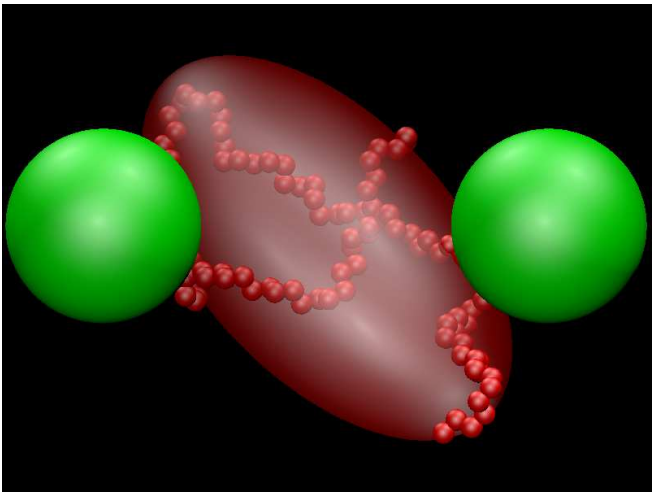}
\caption{
Schematic illustration of model: colloidal particles (large green spheres) 
and polymer coil (smaller red spheres) fit by effective ellipsoid whose
shape fluctuates with the coil.
}\label{snapshot}
\end{figure}

To explore the influence of aspherical polymer conformations and solvent quality 
on the effective interactions induced between colloidal particles by depletion 
of nonadsorbing polymer due to colloid excluded volume, we consider a monodisperse 
suspension of colloidal particles, modeled as hard spheres of radius $R_c$, 
mixed with free polymer coils, modeled as ellipsoids whose shapes fluctuate according 
to the statistics of random walks (Fig.~\ref{snapshot}). 

The strength and range of depletion-induced interactions depend on concentration 
and size of the polymer coils relative to the colloids. Calibrating theoretical
models to experimental systems requires an appropriate measure for the 
effective size of a polymer coil. 
Implementations of the AO model often take the effective radius of a polymer coil 
simply as the rms radius of gyration, defining the polymer-to-colloid size ratio 
as $q\equiv R_g/R_c$. For later reference, we note that, for a given size ratio 
$q_{\scriptscriptstyle\rm RW}$ of a RW polymer in a $\theta$ solvent,
the scaling relations (Sec.~\ref{polymer-ellipsoid}) dictate the 
size ratio $q_{\scriptscriptstyle\rm SAW}$ of a SAW polymer 
of equal molecular weight (same $N$) in a good solvent:
\begin{equation}
q_{\scriptscriptstyle\rm SAW}=C 6^{\nu}(R_c/l)^{2\nu-1}q_{\scriptscriptstyle\rm RW}^{2\nu}~.
\label{q-relation}
\end{equation}
More accurate measures for the effective size of a polymer account for the effect 
on the depletion layer thickness of deformation of a coil near a hard surface.
We first review the simplest case of a polymer near a hard, flat wall, as addressed 
by Asakura and Oosawa~\cite{asakura1954}, and then consider polymers near
hard, spherical colloids.

\begin{figure}[t!]
\includegraphics[width=0.9\columnwidth]{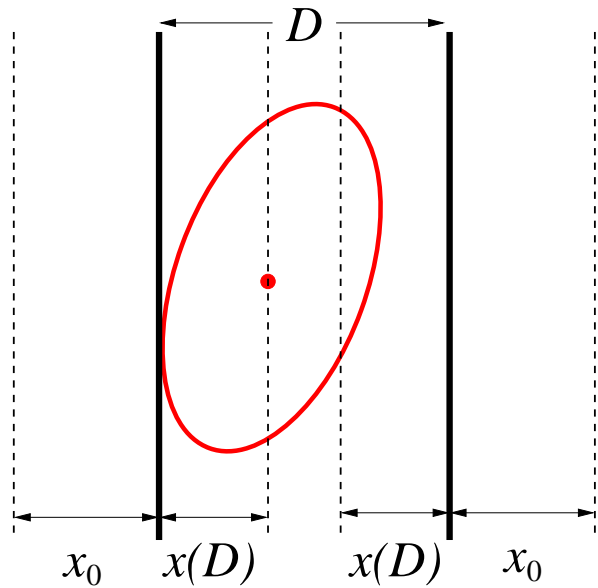}
\caption{
Ellipsoidal polymer, seen in cross-section as an ellipse, inserted into a volume 
containing two infinitely wide parallel hard plates (solid lines) separated by a 
distance $D$. Depletion layers (dashed lines) between and outside the plates 
have thicknesses $x(D)$ and $x_0$, respectively, equal to half the ellipsoid width 
in the direction perpendicular to the walls.
}\label{ellipsoid-plates}
\end{figure}

Consider a solution of $N_p$ polymers in a volume $V$ containing two hard, flat, 
parallel plates of area $A$ separated by a distance $D$ much shorter than the 
lateral extent of the plates ($D\ll\sqrt{A}$). The polymer coils are free to diffuse, 
except for the constraint imposed by the plates. In a closed system, the potential 
of mean force between the plates induced by depletion of polymer from the intervening 
space is defined as the difference between the Helmholtz free energy $F(D)$ at plate
separation $D$ and at infinite separation:
\begin{equation}
v_{\rm mf}(D)\equiv F(D)-F(\infty)~.
\label{vmf-plates}
\end{equation}
In the dilute limit of noninteracting polymers amidst plates separated by distance $D$, 
the free energy is related to the single-polymer partition function ${\cal Z}_1(D)$ via
\begin{equation}
F(D)=-k_BTN_p\ln {\cal Z}_1(D)~.
\label{FD}
\end{equation}
The potential of mean force between plates induced by depletion of 
$N_p$ polymer coils then can be expressed as
\begin{equation}
\beta v_{\rm mf}(D)
=-N_p\ln\left(\frac{{\cal Z}_1(D)}{{\cal Z}_1(\infty)}\right)~.
\label{vmf-plates1}
\end{equation}

Considering that the polymers and plates interact only via excluded-volume interactions,
the single-polymer partition function is simply proportional to the fraction of 
volume available to a polymer:
\begin{equation}
{\cal Z}_1(D)\propto 1-2\,[\la x_0\ra+\la x(D)\ra]\,\frac{A}{V}~,
\label{Z1D}
\end{equation}
where $\la x_0\ra$ and $\la x(D)\ra$ represent average thicknesses of depletion 
layers adjacent to a surface outside of and between the plates, respectively,
and angular brackets denote ensemble averages over depletant conformations
(see Fig.~\ref{ellipsoid-plates}).
For depletants with simple geometrical shapes (e.g., spheres or ellipsoids), 
it can be shown that the depletion layer thickness equals the integrated mean 
curvature $c_d$ of the depletant \cite{oversteegen-roth2005}. For a smooth, 
convex body, $c_d$ is defined as an average over the body's closed surface $S$ 
of the mean curvature,
\begin{equation}
H(\br)=\frac{1}{2}\left(\frac{1}{R_1(\br)}+\frac{1}{R_2(\br)}\right)~,
\label{H}
\end{equation}
where $R_1(\br)$ and $R_2(\br)$ are the local radii of curvature at a point 
${\bf r}$ on the surface. To model depletants that fluctuate in size and shape,
we augment this definition with an average over conformations:
\begin{equation}
c_d\equiv\frac{1}{4\pi}\la\oint_S dS\, H(\br)\ra~.
\label{cd}
\end{equation}
Note that $c_d$ has physical dimensions of length.

From Eq.~(\ref{Z1D}) and the limiting relation
\begin{equation}
x_0=\lim_{D\to\infty} x(D)~,
\label{limit}
\end{equation}
the partition function for a polymer in a system with infinitely separated plates is
\begin{equation}
{\cal Z}_1(\infty)\propto 1-4\,\la x_0\ra\,\frac{A}{V}~.
\label{Z1-infinity}
\end{equation}
Substituting Eqs.~(\ref{Z1D}) and (\ref{Z1-infinity}) into Eq.~(\ref{vmf-plates1}),
and assuming a polymer solution so dilute that the excluded volume is only a small fraction 
of the total volume, 
the PMF between the plates per unit plate area is 
\begin{equation}
w_{\rm mf}(D)\equiv\frac{v_{\rm mf}(D)}{A}
\simeq 2\,\Pi_p\,[\la x(D)\ra-\la x_0\ra]~,
\label{wmf-plates}
\end{equation}
where $\Pi_p=N_pk_BT/V$ is the osmotic pressure of an ideal gas of polymer coils. 
In the limit as the plates come together ($D\to 0$), where $\la x(D)\ra\to 0$, 
the PMF induced by real polymer chains with radius of gyration $R_g$ approaches 
the exact contact value~\cite{asakura1954,doi-edwards1986,lekkerkerker-tuinier2011}
\begin{equation}
|w_{\rm mf}(0)|=2\,\Pi_p\la x_0\ra=\frac{4}{\sqrt{\pi}}\,\Pi_p R_g~.
\label{wmf0-plates}
\end{equation}
Now identifying $\la x_0\ra$ with the integrated mean curvature $c_d$ (averaged over 
conformations) of an uncrowded polymer coil, modeled as a fluctuating ellipsoid
(see Fig.~\ref{ellipsoid-plates}), and defining 
%\textcolor{red}{
$\gamma\equiv c_d/R_{p, {\rm eff}}$ 
%}
as the coefficient of proportionality between $c_d$ and the 
{\it effective} polymer radius $R_{p, {\rm eff}}$, we have
\begin{equation}
c_d=\la x_0\ra=\gamma\,R_{p,{\rm eff}}=\frac{2}{\sqrt{\pi}}\,R_g~,
\label{cd}
\end{equation}
and thus finally,
\begin{equation}
R_{p, {\rm eff}}=\frac{2}{\sqrt{\pi}\,\gamma}\,R_g~.
\label{Rpeff}
\end{equation}
%\textcolor{red}{
We emphasize that $R_{p, {\rm eff}}$ represents the effective radius of a nonadsorbing 
polymer coil in the presence of hard colloidal particles, taking into account both the 
diffuse periphery and aspherical shape of the coil. 
%}
We conclude that an experimental system with nominal polymer-to-colloid size ratio 
$q=R_g/R_c$, defined by reference to the radius of gyration of a real polymer chain, 
should be modeled using an {\it effective} size ratio 
\begin{equation}
q_{\rm eff}\equiv\frac{R_{p, {\rm eff}}}{R_c}=\frac{2}{\sqrt{\pi}\,\gamma}\,
\frac{R_g}{R_c}=\frac{2}{\sqrt{\pi}\,\gamma}\,q~.
\label{qeff}
\end{equation}

For a sphere of fixed radius (AO model), the integrated mean curvature simply equals 
the radius ($\gamma=1$). For a fluctuating ellipsoid, on the other hand, determining $\gamma$ 
is nontrivial. We computed $\gamma$ using two independent, but equivalent, numerical methods.
In the first method, based on Eq.~(\ref{cd}), we numerically integrated the 
mean curvature over the surface of the ellipsoid and averaged over the polymer shape 
probability distribution [Eq.~(\ref{Plambda}) combined with Eq.~(\ref{fi-Eurich-Maass}) 
or (\ref{fi-Sciutto}) for RW or SAW polymers, respectively].
In the second method, we computed the half-width of an ellipsoid in a fixed direction 
and numerically averaged over orientations and the shape probability distribution. 
Both methods involve numerically evaluating a five-dimensional 
integral and both give the same result to within numerical precision. 
From the shape distributions corresponding to a RW polymer ($\theta$ solvent) 
and a SAW polymer (good solvent), we find 
$\gamma_{\scriptscriptstyle\rm RW}=0.93254$ and 
$\gamma_{\scriptscriptstyle\rm SAW}=0.92431$, respectively.

For polymers dispersed in a suspension of hard-sphere colloids, the effective size ratio 
$q_{\rm eff}$ can be objectively defined by equating the free energy cost of inserting
a hard sphere into a solution of polymers, as predicted by polymer field theory,
with the work done to inflate a sphere in the model polymer solution. When applied to 
polymers obeying RW statistics, and presumed spherical in shape and fixed in size
(AO model), this definition yields~\cite{eisenriegler1996,louis2002-jcp1,louis2002-jcp2,aarts2002}
\begin{equation}
q_{\rm eff, \scriptscriptstyle RW-AO}=\left(1+\frac{6}{\sqrt{\pi}}q+3q^2\right)^{1/3}-1~,
\label{qeff-sphere-RW}
\end{equation}
assuming $q_{\rm eff}<1$, such that penetration of a polymer by a colloid can be neglected.
For polymers with aspherical (ellipsoidal), fluctuating shapes dispersed with 
hard-sphere colloids, we modify this definition in the same manner as for polymers 
near a hard, flat wall by incorporating the integrated mean curvature:
\begin{equation}
q_{\rm eff, \scriptscriptstyle RW}=\frac{1}{\gamma_{\scriptscriptstyle\rm RW}}
\left[\left(1+\frac{6}{\sqrt{\pi}}q+3q^2\right)^{1/3}-1\right]~.
\label{qeff-ellipsoid-RW}
\end{equation}
This definition ensures that in the limit $q\to 0$ the model recovers the exact 
contact value of the PMF induced by RW polymers between hard, flat plates.
We emphasize that the $1/\gamma$ adjustment proved essential in our earlier
study~\cite{lim-denton-sm2016} for achieving quantitative agreement with PMF data 
from molecular simulations~\cite{meijer-frenkel1991,meijer-frenkel1994} and 
from experiments on DNA-induced depletion forces~\cite{yodh1998}.

In contrast, for polymers that obey SAW statistics, also presumed spherical and 
of fixed size (AO model), field theory yields an effective size ratio~\cite{schaefer1999,aarts2002}
\begin{equation}
q_{\rm eff, \scriptscriptstyle SAW-AO}=\left(1+C_1q+C_2q^2-C_3q^3\right)^{1/3}-1~,
\label{qeff-sphere-SAW}
\end{equation}
where $C_1\simeq 3.2130$, $C_2\simeq 2.6073$, and $C_3\simeq 0.1197$.
For aspherical (ellipsoidal) SAW polymers, we similarly incorporate the 
integrated mean curvature and define
\begin{equation}
q_{\rm eff, \scriptscriptstyle SAW}=\frac{1}{\gamma_{\scriptscriptstyle\rm SAW}}
\left[\left(1+C_1q+C_2q^2-C_3q^3\right)^{1/3}-1\right]~.
\label{qeff-ellipsoid-SAW}
\end{equation}

\subsection{Potential of Mean Force}\label{simulations}

Adapting Eq.~(\ref{vmf-plates}) from plates to spheres, the potential of mean force
between two colloids in thermal equilibrium at absolute temperature $T$ 
with a solution of nonadsorbing polymers in a closed volume is defined as the change 
in Helmholtz free energy $F(r)$ of the system upon bringing the particles from 
infinite separation to center-to-center separation $r$:
\begin{equation}
v_{\rm mf}(r)=F(r)-F(\infty)~,
\label{veff1}
\end{equation}
since in an isotropic fluid the pair potential depends on only the radial coordinate. 
For a system in chemical equilibrium with a polymer reservoir, the PMF is defined 
as the change in grand potential $\Omega(r)$. In earlier work~\cite{lim-denton-jcp2016},
we applied an alternative (but equivalent) definition, that is more appropriate 
in the nanoparticle limit, in which polymer coils are significantly larger than 
and penetrable by the colloids (nanoparticles).

The free energy varies with colloidal separation due to mechanical work performed 
by the colloids in changing the excluded volume of the polymer with osmotic pressure 
$\Pi_p=n_p k_BT$, assuming a dilute (ideal gas) polymer solution of mean density $n_p$.
In the AO model, this work is easily evaluated:
\begin{equation}
v_{\rm mf}(r)=-\Pi_p\int_{\infty}^r dr'\, A_{\rm ov}(r')=-\Pi_p V_{\rm ov}(r)~,
\label{veff2}
\end{equation}
where $A_{\rm ov}(r)$ and $V_{\rm ov}(r)$ are the cross-sectional area and volume, 
respectively, of the overlap region of the two excluded-volume shells and we choose 
$F(\infty)=0$. It should be noted that, when used to model thermodynamic phase behavior,
Eq.~(\ref{veff2}) must be corrected for triplet overlaps at size ratios above
$q\simeq 0.1547$, especially away from the dilute colloid concentration limit.
For spherical colloids and spherical polymers of fixed radius $R_p$, 
the convex-lens-shaped pair overlap region, defined by the intersection of 
two spherical excluded-volume shells, has volume 
\begin{equation}
V_{\rm ov}(r)=
\frac{\displaystyle 4\pi}{\displaystyle 3}
\left[\left(R_c+R_p\right)^3
-\frac{\displaystyle 3r}{\displaystyle 4}(R_c+R_p)^2+
\frac{\displaystyle r^3}{\displaystyle 16}\right]
\label{AO}
\end{equation}
for $2R_c<r<2(R_c+R_p)$ (otherwise zero). 
%Added:
%\textcolor{red}{
Equations~(\ref{veff2})-(\ref{AO}) express the conventional AO potential.
%}
%

In the case of aspherical depletants,
this simple geometric approach can be adapted by calculating an {\it average} of 
the overlap volume $\la V_{\rm ov}(r)\ra$ over an ensemble of polymer conformations 
(orientations and shapes). From a large sample of randomly generated conformations 
(microstates), $\la V_{\rm ov}(r)\ra$ equals the sampled volume times the fraction 
of microstates in which a depletant overlaps both colloids.

%Added:
%\textcolor{red}{
Two limitations of our modeling approach are important to note. First, the 
coarse-grained model of polymer coils necessarily neglects coil shapes that 
deviate from ellipsoidal, which may affect how nonadsorbing polymers interact 
with hard colloidal surfaces. Second, although neighboring coils in a 
$\theta$ solvent do not influence each other's shape distribution, since 
polymer segments are effectively noninteracting, excluded-volume interactions 
between segments in a good solvent can lead to correlations between shapes 
of neighboring coils. Such correlations may affect the strength and range 
of polymer depletion-induced interactions, especially in semi-dilute or 
concentrated polymer solutions. The present modeling approach, 
which describes only the PMF induced by independent ellipsoidal polymer coils, 
neglects such effects. In Sec.~\ref{results}, we discuss implications and
potential remedies of these limitations of our approach.
%}
%

\section{Monte Carlo Simulations}\label{simulations}

To compute the potential of mean force between colloids induced by depletion 
of nonadsorbing polymers that fluctuate in size and shape according to 
either RW chain statistics ($\theta$ solvent) or SAW statistics (good solvent),
we used Monte Carlo (MC) simulation methods~\cite{frenkel-smit2001}.
Applying Eq.~(\ref{veff2}), we determined the average overlap volume 
$\la V_{\rm ov}(r)\ra$ by placing two hard-sphere colloids in a rectangular 
parallelepiped simulation box at center-to-center separation $r$, inserting a 
polymer ellipsoid at a random position with random orientation and shape 
governed by the appropriate gyration tensor eigenvalue probability distribution
[Eq.~(\ref{fi-Eurich-Maass}) or (\ref{fi-Sciutto})], and counting the fraction 
of double overlaps, i.e., insertions leading to an overlap of the ellipsoid
with both spheres. 

%Added:
%\textcolor{red}{
As noted in Sec.~\ref{models}, since our model constrains polymer coils 
to have only ellipsoidal shapes, our approach, although it captures the
gross shapes of polymers, neglects any influence of non-ellipsoidal conformations 
on polymer-colloid interactions. Furthermore, since we insert polymer coils 
only one at a time, our approach, when applied to SAW polymers in good solvents,
is limited to dilute polymer solutions, since it neglects possible correlations 
between shapes of neighboring, interacting coils. 
%}
%

To randomly sample polymer conformations, we implemented a variation of the 
Metropolis algorithm~\cite{frenkel-smit2001}.
Trial changes in orientation and shape of a polymer ellipsoid were coupled with insertions.
To uniformly sample orientations, specified by a unit vector ${\bf u}$ aligned 
with the longest axis of the ellipsoid, we generated a new (trial) unit vector,
${\bf u}'=({\bf u}+\tau{\bf v})/|{\bf u}+\tau{\bf v}|$, where ${\bf v}$ is a 
randomly oriented unit vector and $\tau$ is a tolerance~\cite{frenkel-smit2001}.
A trial change in shape from one set of gyration tensor eigenvalues 
$\lambda\equiv\{\lambda_i\}$ to a new set 
$\lambda'\equiv\{\lambda_i'=\lambda_i+\Delta\lambda_i\}$ with tolerances 
$\Delta\lambda_i$ ($i=1,\ldots,3$) implies a change $\Delta F_c$ in the coil's 
internal free energy~\cite{doi-edwards1986}, $F_c=-k_BT\ln P(\lambda)$,
where $P(\lambda)$ is the polymer shape distribution [Eq.~(\ref{Plambda}) with 
Eq.~(\ref{fi-Eurich-Maass}) or (\ref{fi-Sciutto})].
A trial conformation was rejected if the inserted polymer ellipsoid overlapped 
either colloidal sphere. Otherwise, it was accepted with probability
\begin{equation}
{\cal P}(\lambda\to\lambda')=
\min\left\{e^{-\beta \Delta F_c},~1\right\} 
=\min\left\{\frac{P(\lambda')}{P(\lambda)},~1\right\}~.
\label{shape-variation}
\end{equation}
If the trial conformation was accepted, the ellipsoid's orientation and shape 
were updated and the double-overlap counter was incremented.
Limiting our study to dilute solutions, we inserted polymers one at a time,
thus neglecting polymer-polymer interactions.
To diagnose overlap of a colloid and a polymer, we computed the shortest distance 
between a point (sphere center) and the ellipsoid surface, requiring
evaluating the roots of a 6th-order polynomial~\cite{heckbert1994}. 
This sampling method yields the average volume of the polymer depletion region 
surrounding two colloidal spheres and hence, from Eq.~(\ref{veff2}), the PMF.

\section{Results and Discussion}\label{results}
 
To compare potentials of mean force between colloidal hard spheres induced by 
depletion of nonadsorbing polymer in $\theta$ and good solvents, we implemented 
the ellipsoidal polymer model described in Sec.~\ref{models} and performed 
a series of Monte Carlo simulations. 
Using the polymer trial insertion method outlined in Sec.~\ref{simulations},
we computed the PMF over a range of colloid separations. The side lengths of 
the rectangular parallelepiped simulation box were set small enough to maximize 
the acceptance ratio, while large enough to avoid interaction of a polymer 
with periodic images of the colloids. Tolerances for polymer trial moves, 
optimized by trial and error, were fixed at $\tau=0.001$ for rotations and 
$\Delta\lambda_1=0.01$, $\Delta\lambda_2=0.003$, and $\Delta\lambda_3=0.001$
for shape changes. For a given colloid pair separation, we performed $10^6$ 
independent trial polymer insertions, and then computed statistical uncertainties 
(error bars) as standard deviations from five independent runs for a total 
of $5\times 10^6$ trial insertions.

To validate our methods, we first implemented the original AO model of spherical polymers 
of fixed size and confirmed that our algorithm reproduces the exact PMF predicted by
Eqs.~(\ref{veff2}) and (\ref{AO}). We then proceeded to simulate the ellipsoidal 
polymer model for polymers whose sizes and shapes are governed by RW and SAW chain statistics,
corresponding to polymers in $\theta$ and good solvents, respectively.
To compare depletion of polymers of equal segment number in different solvents, 
we converted polymer-to-colloid size ratios between RW and SAW statistics using 
Eq.~(\ref{q-relation}). This conversion requires specifying the ratio of the colloid
radius $R_c$ to the polymer segment length $l$. To make potential contact with experiments, 
we chose typical values of $R_c=100$ nm and $l=0.76$ nm, corresponding 
to polyethylene glycol (PEG) in water~\cite{LEE20081590}.

\begin{table}[h]
\caption{Polymer-to-colloid size ratios for MC simulations to compute PMF 
(Figs.~\ref{fig-pmf1}, \ref{fig-pmf2}) between hard-sphere colloids 
induced by depletion of nonadsorbing, linear polymer coils 
modeled as non-self-avoiding random walks (RW) or self-avoiding walks (SAW) in 
$\theta$ and good solvents, respectively. Tabulated from left to right are bare and 
effective size ratios of ellipsoidal RW polymers, $q_{\scriptscriptstyle\rm RW}$ and 
$q_{\rm eff, \scriptscriptstyle RW}$, bare and effective size ratios of ellipsoidal SAW 
polymers, $q_{\scriptscriptstyle\rm SAW}$ and $q_{\rm eff, \scriptscriptstyle SAW}$,
effective size ratio of spherical SAW polymers in the AO model, 
$q_{\rm eff, \scriptscriptstyle SAW-AO}$, and corresponding number of segments $N$.}
\vspace*{0.2cm}
\centering
\begin{tabular}{|c|c|c|c|c|c|}
\hline
$q_{\scriptscriptstyle\rm RW}$ & 
$q_{\rm eff, \scriptscriptstyle RW}$ & 
$q_{\scriptscriptstyle\rm SAW}$ & 
$q_{\rm eff, \scriptscriptstyle SAW}$ & 
$q_{\rm eff, \scriptscriptstyle SAW-AO}$ & 
$N$ \\
\hline
\hline
0.1&0.11821&0.19909&0.21989&0.20325&1040 \\
\hline
0.2&0.23138&0.44984&0.47186&0.43614&4155 \\
\hline
0.3&0.34018&0.72467&0.72416&0.66935&9350 \\
\hline
0.4&0.44518&1.01641&0.97116&0.89765&16620 \\
\hline
\end{tabular}
\label{table3}
\end{table}

\begin{center}
\begin{figure}
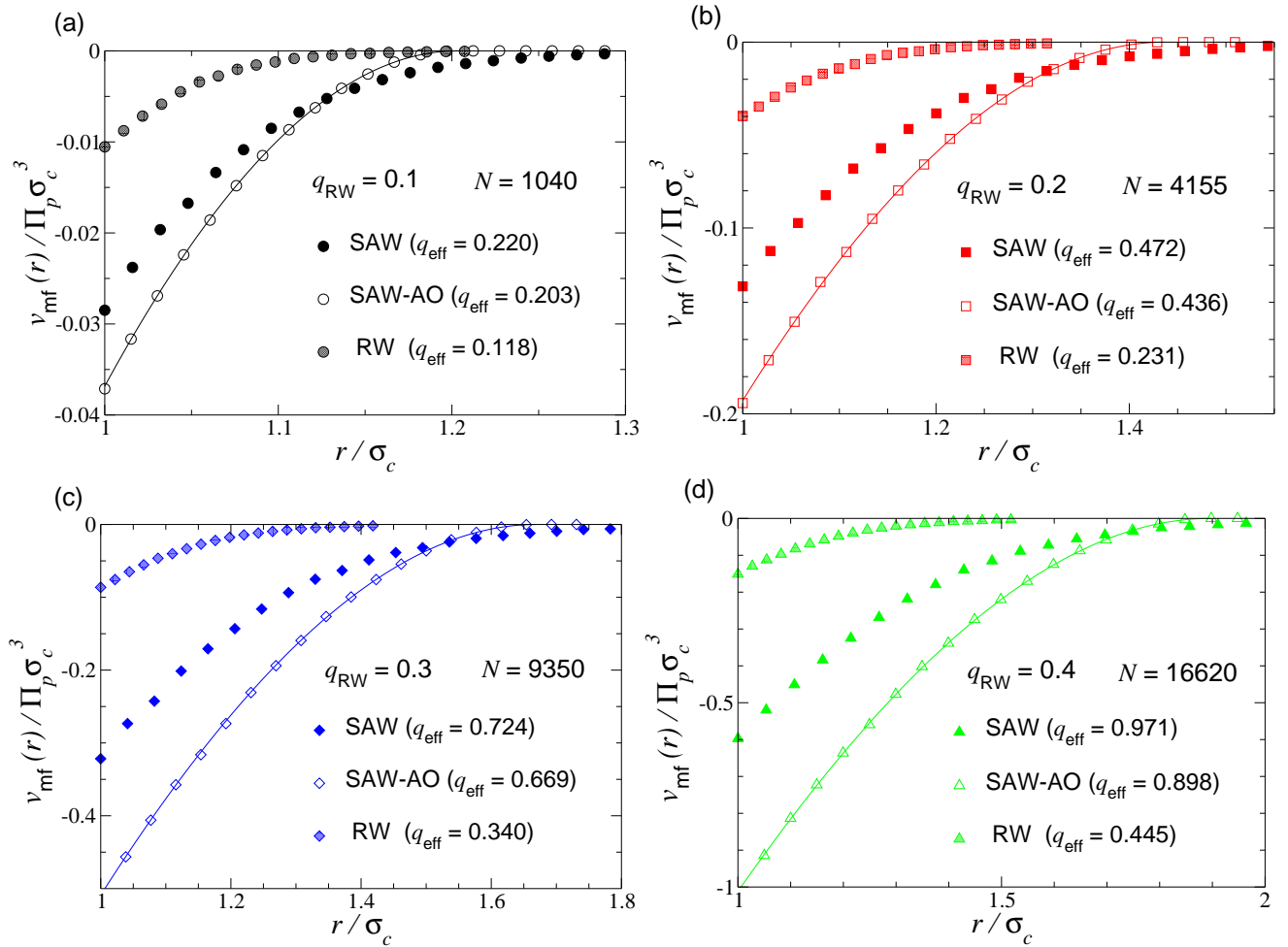

\includegraphics[width=0.48\columnwidth]{pomf-saw.N1040.eps}
\includegraphics[width=0.48\columnwidth]{pomf-saw.N4155.eps}
\\[0.2ex]
\includegraphics[width=0.48\columnwidth]{pomf-saw.N9350.eps}
\includegraphics[width=0.48\columnwidth]{pomf-saw.N16620.eps}
\vspace*{-0.2cm}
\caption{
Potential of mean force $v_{\rm mf}(r)$ (units of $\Pi_p\sigma_c^3$) between 
hard-sphere colloids vs.~separation $r$ (units of colloid diameter $\sigma_c$)
induced by nonadsorbing polymers for random-walk polymer-to-colloid size ratio 
$q_{\scriptscriptstyle\rm RW}=0.1$ (a), $0.2$ (b), $0.3$ (c), and $0.4$ (d). 
For corresponding effective size ratios $q_{\rm eff}$, simulation data are shown 
for self-avoiding walk (SAW) polymers in a good solvent (solid symbols) and 
non-self-avoiding random walk (RW) polymers in a $\theta$ solvent (lightly-shaded symbols)
-- both modeled as fluctuating ellipsoids -- and for AO polymers in a good solvent 
(open symbols), modeled as rigid spheres. Curves represent analytical expression 
for AO potential [Eqs.~(\ref{veff2}) and (\ref{AO})]. 
Statistical error bars are smaller than symbol sizes.
}\label{fig-pmf1}
\end{figure}
\end{center}

Results of our calculations for the PMF are presented in Fig.~\ref{fig-pmf1}
over a range of size ratios. 
%Added:
%\textcolor{red}{
Note that the vertical axis is scaled by the polymer osmotic pressure $\Pi_p$,
rendering the plotted PMF independent of polymer concentration. As discussed
above in Secs.~\ref{models} and \ref{simulations}, however, the results shown
for SAW polymers are physically meaningful only for dilute polymer solutions.
%}
%
For each bare size ratio of RW polymers in the series
$q_{\scriptscriptstyle\rm RW}=0.1, 0.2, 0.3, 0.4$, we calculated the corresponding
bare size ratio of SAW polymers $q_{\scriptscriptstyle\rm SAW}$ of equal $N$ from 
Eq.~(\ref{q-relation}). We then ran simulations for the effective size ratios of 
ellipsoidal RW and SAW polymers, $q_{\rm eff, \scriptscriptstyle RW}$ and 
$q_{\rm eff, \scriptscriptstyle SAW}$, calculated from Eqs.~(\ref{qeff-ellipsoid-RW}) 
and (\ref{qeff-ellipsoid-SAW}), respectively. For comparison, we also simulated 
spherical SAW polymers (AO model) with effective size ratio 
$q_{\rm eff, \scriptscriptstyle SAW-AO}$, calculated from Eq.~(\ref{qeff-sphere-SAW}).
The system parameters are tabulated in Table~\ref{table3}.

\onecolumngrid

\twocolumngrid
\begin{center}
\begin{figure}
\includegraphics[width=\columnwidth]{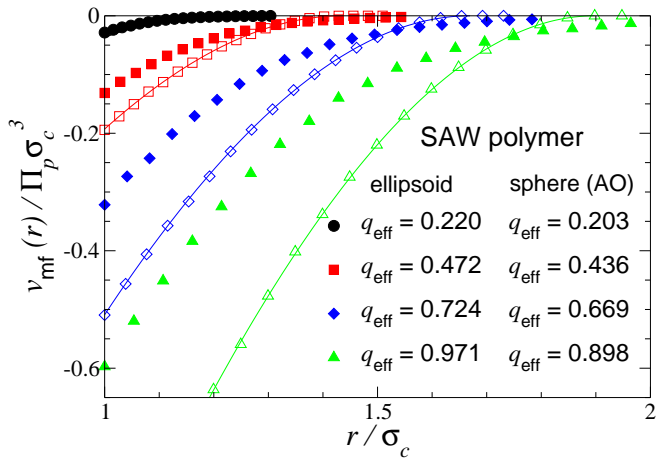}
\vspace*{-0.2cm}
\caption{
Potential of mean force $v_{\rm mf}(r)$ (units of $\Pi_p\sigma_c^3$) between hard-sphere 
colloids vs.~separation $r$ (units of colloid diameter $\sigma_c$) induced by nonadsorbing 
SAW polymers in a good solvent. Over a range of effective polymer-to-colloid 
size ratios $q_{\rm eff}$, simulation data are shown for fluctuating ellipsoidal polymers 
(solid symbols) and fixed-size spherical (AO model) polymers (open symbols). 
Curves represent analytical expression for AO potential [Eqs.~(\ref{veff2}) and (\ref{AO})].
Statistical error bars are smaller than symbol sizes.
}\label{fig-pmf2}
\end{figure}
\end{center}

In Fig.~\ref{fig-pmf1}, the solid symbols represent PMF data for polymers in
a good solvent, modeled as fluctuating ellipsoids obeying SAW chain statistics
[Eq.~(\ref{fi-Eurich-Maass})]. The lightly-shaded symbols represent PMF data for 
polymers in a $\theta$ solvent, modeled as fluctuating ellipsoids obeying RW chain 
statistics [Eq.~(\ref{fi-Sciutto})]. For comparison, the open symbols represent 
PMF data for polymers in a good solvent, modeled as spheres of fixed size (AO model).
Since in the dilute limit, the PMF is proportional to the polymer osmotic pressure,
we plot the dimensionless quantity of $v_{\rm mf}(r)$ scaled by $\Pi_p$ times 
the cube of the colloid diameter $\sigma_c=2R_c$. Note that negative values of
$v_{\rm mf}(r)$ imply an attractive pair interaction between colloids.
To within statistical uncertainty, our data for the AO model are perfectly fit
by the analytical expressions of Eqs.~(\ref{veff2}) and (\ref{AO}) (curves).

Having validated our methods, we now compare our PMF data for polymer coils 
modeled as fluctuating ellipsoids (solid symbols in Fig.~\ref{fig-pmf1}) 
with corresponding data for coils of equal segment number modeled as spheres of 
fixed size (open symbols in Fig.~\ref{fig-pmf1}), both obeying SAW chain statistics.
Evidently, fluctuating ellipsoid polymers induce generally weaker PMFs than 
fixed-sphere polymers (AO model). For ellipsoidal polymers, the contact value 
$v_{\rm mf}(\sigma_c)$ is consistently lesser in magnitude, while the range of 
$v_{\rm mf}(r)$ is consistently longer than for spherical polymers. 
Furthermore, deviations between the PMFs from the ellipsoidal and spherical 
polymer models grow with increasing size ratio. 

These results may appear surprising, considering that depletion-induced attraction 
in the AO model strengthens with increasing size ratio, and given that
the effective size ratio in the ellipsoidal polymer model [Eq.~(\ref{qeff-ellipsoid-SAW})]
exceeds that in the spherical polymer model [Eq.~(\ref{qeff-sphere-SAW})] 
(the integrated mean curvature being smaller for an ellipsoid than for a sphere 
of equal radius of gyration). Nevertheless, these trends are quite consistent with 
the extra conformational freedom of fluctuating ellipsoids to elongate to lengths 
beyond their mean diameter and to deform to avoid hard surfaces. 
In previous work~\cite{lim-denton-sm2016}, we showed that the fluctuating ellipsoid 
polymer model, when implemented with the appropriate effective size ratio, 
nearly exactly reproduces the PMF computed from ``lattice polymer" simulations
of RW polymers whose segments are confined to the sites of a cubic 
lattice~\cite{meijer-frenkel1991,meijer-frenkel1994}.

Next, we compare our results for PMFs induced by depletion of polymer coils, modeled as 
fluctuating ellipsoids, that obey either SAW or RW chain statistics.
From Fig.~\ref{fig-pmf1}, we see that SAW polymers in a good solvent (solid symbols)
induce PMFs that are significantly stronger -- both greater in magnitude and longer 
in range -- than RW polymers of equal segment number in a $\theta$ solvent 
(lightly-shaded symbols). Qualitatively, this trend is consistent with the more 
extended conformations and correspondingly larger effective radius of gyration
of SAW polymers compared with RW polymers of the same contour length. Quantitatively,
it is interesting that, at least for the system parameters considered here, the 
attractive well of the PMF induced by SAW polymers is roughly three times deeper 
than that of the PMF induced by RW polymers.

Figure~\ref{fig-pmf2} collects data from Fig.~\ref{fig-pmf1} in one plot 
to summarize the dependence on polymer-to-colloid size ratio of the PMF induced by 
SAW polymers. With increasing effective size ratio, the depth and range of $v_{\rm mf}(r)$
both steadily grow. As noted above, the fluctuating ellipsoid polymer model 
predicts a generally weaker, but longer-ranged, PMF compared with the AO model. 
Although we are not aware of lattice-polymer simulations of colloids dispersed 
in SAW polymer solutions, our predictions could be tested against such 
molecular-scale simulations.

%Added:
%\textcolor{red}{
As mentioned in Secs.~\ref{models} and \ref{simulations}, our modeling approach
is based on two main approximations. First, it neglects non-ellipsoidal shapes 
of nonadsorbing polymer coils interacting with colloidal surfaces and, second, 
it neglects interactions and associated shape correlations between neighboring coils.
The accuracy of the first approximation can be quantified by comparing predictions 
of the coarse-grained polymer model with those of molecular-scale polymer models. 
In our previous study of depletion interactions induced by polymers in a $\theta$ 
solvent~\cite{lim-denton-sm2016}, we compared our PMF results against simulations 
of a model of RW polymers on a lattice, demonstrating remarkable accuracy of the 
coarse-grained polymer model. A similar comparison for SAW polymers in a good solvent 
could further assess the accuracy of the coarse-grained model.
%}

%\textcolor{red}{
The second approximation is well justified for polymers in $\theta$ solvents,
except to the extent that the $\theta$ temperature may vary with polymer concentration.
For SAW polymers in good solvents, however, neglecting interactions between neighboring 
coils strictly limits application of our approach to dilute polymer solutions. 
Consequently, the model may not accurately describe the influence of depletion forces
on thermodynamic properties, including bulk phase separation in concentrated mixtures 
of colloids and polymers in good solvents. Extending the model beyond the dilute regime 
would require incorporating the influence on polymer conformations of interactions and 
correlations between segments of different coils. This extension could be achieved, 
for example, by simulating a concentrated solution of a few explicit SAW interacting polymers,
computing their gyration tensors, and fitting the eigenvalue probability distributions
over a range of polymer concentrations.
%}
%

\section{Summary and Conclusions}\label{conclusions}

In summary, we have implemented Monte Carlo simulation methods for computing the 
potential of mean force between hard-sphere colloids induced by nonadsorbing polymer 
coils dispersed in good and $\theta$ solvents. For computational efficiency, 
we modeled the polymer coils as general ellipsoids, whose conformations 
(size and shape) fluctuate according to statistics of either self-avoiding walks 
(good solvent) or non-self-avoiding random walks ($\theta$ solvent). The principal
radii of the equivalent ellipsoid representing a polymer coil were determined from 
accurate fits to probability distributions for the eigenvalues of the gyration tensor
of a SAW or RW.

In the colloid limit, in which the polymer radius of gyration is smaller than the 
colloid radius, we determined the PMF by computing the average volume of the polymer 
depletion region surrounding a pair of colloids using a polymer insertion algorithm.
Because polymer conformational distributions vary with solvent quality, the average
depletion volume, and therefore the PMF, differ between good and $\theta$ solvents.
Our results demonstrate that the dependence of the PMF on polymer shape and 
solvent quality can be quite significant. 

Comparing the ellipsoidal and spherical (AO) models of polymers in good solvents,
we showed that the former model yields a generally weaker PMF, which we attribute 
to the freedom allowed by the fluctuating ellipsoid model for a polymer coil to 
adapt its shape to a crowded environment. This finding is consistent with conclusions 
from our previous studies of depletion interactions in colloid-polymer mixtures 
dispersed in $\theta$ solvents~\cite{lim-denton-sm2016,lim-denton-jcp2016}.

Comparing the ellipsoidal model of SAW and RW polymers in good and $\theta$ solvents,
respectively, we showed that polymers in good solvents induce considerably stronger
PMFs than polymers of the same number of segments in $\theta$ solvents.
This trend is explained by the more extended conformations of SAW polymers, 
which tend to enlarge the depletion region around colloids.
%Added:
%\textcolor{red}{
The coarse-grained model of depletion interactions induced by nonadsorbing polymers
could be further tested against experiments and molecular-scale models of polymers, 
especially for SAW coils dispersed in good solvents. 
%}
%Deleted:
%The predictions of our modeling can be tested by experiments and molecular-scale 
%simulations. 

Our work suggests the possibility of tuning effective depletion-induced interactions 
between colloids, and thereby thermodynamic phase behavior of colloid-polymer mixtures,
by varying solvent quality, e.g., by selecting particular polymer-solvent combinations 
or by changing temperature and cosolvent concentration for a given combination.
Our modeling approach, which focuses on the geometry of the depletant, may also have 
relevance for effective interactions induced by other types of soft depletants that 
can vary in size and shape, such as vesicles~\cite{taniguchi2008} and 
microgels~\cite{schurtenberger2014,schurtenberger2018}.

Future work could explore the influence of solvent quality on depletion-induced 
interactions in the protein limit, in which polymers are large enough to be 
penetrated by colloids. As shown in our previous studies of polymer-nanoparticle 
mixtures~\cite{lim-denton-jcp2016,davis-denton2018}, accurate modeling would require 
a reliable expression (from polymer field theory) for the penetration free energy.
In this regime, our geometric approach may yield insights complementary to those 
provided by field theories~\cite{eisenriegler1996,hanke1999,eisenriegler2003,odijk2000,
forsman2010,forsman2012,forsman2014,ganesan2005},
integral-equation theories~\cite{chatterjee1998a,chatterjee1998b,chatterjee1999,
schweizer2002,moncho-jorda2003}, and density-functional theories~\cite{leiderer1999,
schmidt-fuchs2002,mukherjee2004,forsman2008,forsman2009}.
%Added:
%\textcolor{red}{
The model also could be extended to concentrated polymer solutions by incorporating 
interactions and correlations between segments within different coils and 
correspondingly modifying the polymer shape distributions, accounting for the 
possible role of polymer density fluctuations near polymer-solvent demixing 
critical points~\cite{chatterjee1999}. 
%}
%
%Deleted:
%and could be extended into the semi-dilute polymer regime, requiring modeling 
%excluded-volume interactions between polymers, and considering the possible role of 
%polymer density fluctuations near polymer-solvent demixing critical points~\cite{chatterjee1999}.
%
%Added:
%\textcolor{red}{
Finally, our approach could be generalized to model depletion interactions between 
aspherical hard colloids~\cite{glotzer2016}, such as rods or platelets.
%}
%Deleted:
%our approach also could be applied to aspherical colloids (e.g., rods or platelets).
%and could be extended into the semi-dilute polymer regime, requiring modeling 
%excluded-volume interactions between polymers, and considering the possible role of 
%polymer density fluctuations near polymer-solvent demixing critical points~\cite{chatterjee1999}.

\vspace*{1cm}
\acknowledgments
This work was supported by the National Science Foundation (Grant No.~DMR-1928073).
We thank Wei Kang Lim for important contributions to coding of the simulations
and Sergio J.~Sciutto for helpful correspondence regarding the SAW shape distribution.

\vspace*{1cm}
\begin{center}
{\bf DATA AVAILABILITY}
\end{center}

The data that support the findings of this study are available from the corresponding author 
upon reasonable request.

%%%%%%%%%%%%%%%%%%%%%%%%%%%%%%%%%%%%%%%%%%%%%%%%%%%%%%%%%%%%%%%%%%%%%%%%%%%
%%%%%%%                       REFERENCES
%%%%%%%%%%%%%%%%%%%%%%%%%%%%%%%%%%%%%%%%%%%%%%%%%%%%%%%%%%%%%%%%%%%%%%%%%%%

\bibliography{depletion,depletion-sm,solvent}

\end{document}